\def\bar{\begin{eqnarray}}
\def\ear{\end{eqnarray}}
\def\bb{\bibitem}
\def\eqi{\begin{equation}}
\def\eqf{\end{equation}}
\def\eqia{\begin{eqnarray}}
\def\eqfa{\end{eqnarray}}
\def\rp#1#2{{#1\over#2}}

\def\lb#1{\label{#1}}

\def\virg#1{``#1''}

\def\oc2{$\mathcal{O}(c^{-2})$}

\documentclass{PoS}

\title{Long-Range Models of Modified Gravity and Their Agreement with Solar System and Double Pulsar Data}

\ShortTitle{Long-Range Models of Modified Gravity and Their Agreement with Observations}

\author{\speaker{Lorenzo Iorio}\\
        INFN-Sezione di Pisa\\
        E-mail: \email{lorenzo.iorio@libero.it}}

\abstract{Many long-range modifications of the Newtonian/Einsteinian standard laws of gravity have been proposed in the recent past to explain various celestial phenomena occurring at different scales ranging from solar system to the entire universe. The most famous ones are the so-called Pioneer anomaly, {i.e.} a still unexplained acceleration detected in the telemetry of the Pioneer 10/11 spacecraft after they passed the 20 AU threshold in the solar system, the non-Keplerian profiles of the velocity rotation curves of several galaxies and the cosmic acceleration. We use the latest observational determinations of the planetary motions in the solar system and in the double pulsar system to put constraints on such models independently of the phenomena for which they were originally proposed. We also deal with the recently detected anomalous perihelion precession of Saturn and discuss the possibility that it can be explained by some of the aforementioned models of modified gravity.}

\FullConference{5th International School on Field Theory and Gravitation,\\
		 April 20 - 24 2009\\
		 Cuiab$\acute{a}$ city, Brazil}

\begin{document}

\section{Motivations for Long-Range Modifications of Gravity}
Historically, the first attempt to modify the laws of gravity commonly accepted at that time was due to Laplace \cite{Lap}  who, in 1805, tried to add a velocity-dependent term
to the standard inverse-square law of Newton to account for the finite velocity of propagation  of gravity.  But this work did not find echo
practically until the surroundings of 1880, when a series of works to estimate
the gravitational finite propagation speed began.  Such attempts to find deviations from the Newtonian inverse-square law
of gravitation were performed to explain the anomalous secular precession of Mercury's
perihelion, discovered by Le Verrier \cite{lev}, without invoking undetected (baryonic) matter like the hypothesized planet Vulcan:  for example, Hall \cite{hal} noted that he could account for Mercury's precession if
the law of gravity, instead of falling off as $1/r^2$, actually falls of as $1/r^k$ with $k = 2.00000016$.
However, such an idea was not found to be very appealing, since it conflicts with basic
conservation laws, e.g., Gauss Law, unless one also postulates a correspondingly modified
metric for space. Other historical attempts to modify Newton's law of gravitation to
account for the Mercury's perihelion behavior yielded velocity-dependent additional terms:
for a review of them see Ref.~\cite{gine} and references therein. Such attempts practically ceased after
the successful explanation of the perihelion rate of Mercury by Einstein \cite{ein} in terms of his tensorial general
theory of relativity: an exception is represented by  Manev \cite{man} who, with a $1/r^2$ correction to the
Newtonian potential, was able to reproduce the anomalous apsidal precession of Mercury.

Moving to more recent times,
 in the modern framework of the challenge of unifying gravity with the other three fundamental interactions of Nature, it was realized that possible
new phenomena could show up just as deviations from the Newtonian inverse-square law of gravitation \cite{nuovo1,nuovo2}.  In general, they would occur at submillimeter length scales.
Concerning general relativity, traditionally, corrections to it, in the form of modifications of the Einstein-Hilbert action in order
to include higher-order curvature invariants with respect
to the Ricci scalar, were considered to be important only at scales close to the Planck one and, consequently, in the early universe or near black hole singularities\footnote{It was shown that in such ways the non-renormalizability of general relativity  became more tractable in the context of quantum field theory.}; see, e.g., Ref~\cite{Prima} and references therein. It was not
expected that such corrections could affect the gravitational
phenomenology at low energies, and consequently
at  larger, macroscopic scales.

Thus, why considering modifications of the standard laws of gravity at large, astronomical, astrophysical and cosmological as well, as done in recent years? To try to accommodate  some recently observed phenomena, occurring at very different scales ranging from solar system to cosmological distances, which, at present, have not yet found fully satisfactorily mundane explanations in terms of conventional physics, gravitational or not \cite{lamm,Berto,megaberto}.  Below we list just some of such  anomalous effects.
\begin{itemize}
\item The flyby anomaly. It consists  of unexplained changes of the asymptotic outgoing velocities of some spacecraft
(Galileo, NEAR, Cassini, and MESSENGER) that occurred at their closest
approaches with the Earth \cite{flyby1,flyby2}. Is it due to  { conventional non-conservative effects}, or are { modifications of the laws of gravity} responsible of it?
\item The {anomalous} {perihelion} precession of {Saturn} \cite{IorSat} detected by processing almost one century of planetary observations with the inclusion of the latest radiometric data of the Cassini spacecraft \cite{Pit08}: is it a {{data processing artifact}} or a genuine {{ physical effect}}?
\item The {Pioneer anomaly}. It is an unexplained {acceleration}  ${ A_{\rm Pio} = (8.74\pm 1.33)\times 10^{-10}} $ {  m s$^{-2}$  } approximately directed towards the Sun affecting the {Pioneer 10/11} spacecraft after they passed the {20 AU} threshold in the {solar system}  \cite{Pio1,Pio2}. Is it induced by some {mundane non-conservative effects}, or is it a sign of { modifications of the laws of gravity}?
\item The dark matter issue.
 In many astrophysical systems like, e.g., spiral galaxies and clusters of galaxies a discrepancy between the observed kinematics of their exterior parts and the predicted one on the basis of the Newtonian dynamics and the matter detected from the emitted electromagnetic radiation (visible stars and gas clouds) was present  since the pioneering studies by Zwicky \cite{Zwi} on the Coma cluster, and by Kahn and Woltjer \cite{Kahn}, Bosma \cite{Bos}, and Rubin and coworkers \cite{Rub,Rub83} on individual  galaxies. More precisely, such an effect shows up in the galactic  velocity rotation curves \cite{flatt} whose typical pattern after a few kpc from the center differs from the Keplerian $1/\sqrt{r}$ fall-off expected from the usual dynamics applied to the electromagnetically-observed matter.  Does the cause of such a phenomenology reside in the action of still undetected (non-baryonic) dark matter  whose dynamics is governed by the usual Newtonian laws of gravitation, or  have they to be modified?
\item The dark energy issue. In recent years, an increasing amount of observational evidence has accumulated pointing toward   the fact that the universe has entered a phase of accelerating expansion. Some of such observations are of direct, geometrical nature: standard candles like the supernov{\ae} SnIa \cite{sna1,sna2}, gamma ray bursts \cite{gamma} and standard rulers like the CMB sound horizon \cite{CMB1,CMB2}. Other ones are of dynamical nature like the rate of growth of cosmological perturbations \cite{grow} probed by the redshift distortion factor or by weak lensing \cite{Fu}.  All these observations are converging towards a confirmation of the accelerating expansion of the universe, assumed homogeneous. They are successfully fitted by  the simplest cosmological model predicting accelerating cosmic expansion: its
    ingredients are the the assumptions of flatness, validity of general relativity, the presence of a cosmological constant $\Lambda$, identified with an unknown and still directly undetected form of energy (named dark energy for these reasons), and Cold Dark Matter ($\Lambda$CDM). However, for some puzzles of the $\Lambda$CDM cosmology, see Ref.~\cite{Peri}. Contrary to the $\Lambda$CDM paradigm, is it possible to accommodate the aforementioned observations by invoking modifications of the standard laws of gravity?
\end{itemize}
Some of the models of modified gravity that have been proposed to address the aforementioned phenomenology are listed below.
\begin{itemize}
  \item Dvali-Gabadadze-Porrati ({DGP})  braneworld model \cite{DGP}. In it our universe is a
(3+1) space-time brane embedded in a five-dimensional Minkowskian bulk. All the
particles and fields of our experience are constrained to remain on the brane apart
from gravity which is free to explore the empty bulk. Beyond a certain threshold
$r_0$, which is a free-parameter of the theory and is fixed by observations to about 5 Gpc, gravity experiences strong modifications with respect to the usual
four-dimensional Newton-Einstein picture: they allow to explain the observed acceleration
of the expansion of the Universe without resorting to the concept of
dark energy.
  \item MOdified Gravity ({MOG}) by Moffat \cite{MOG}.  It is a fully covariant theory of gravity which is based on
the existence of a massive vector field coupled universally
to matter. MOG yields a Yukawa-like modification
of gravity with three constants which, in the most general
case, are running; they are present in the theory's action
as scalar fields which represent the gravitational constant,
the vector field coupling constant, and the vector field
mass. It has used to successfully describe various observational phenomena
on astrophysical and cosmological scales without
resorting to dark matter \cite{Mof2}.
  \item MOdified Newtonian Dynamcs ({MOND}) \cite{Mil83a,Mil83b,Mil83c}. It is a non-linear theory of gravity which predicts departures from the standard Newtonian inverse-square law at a characteristic acceleration scale \cite{Bege} $A_0=1.27\times 10^{-10}$ m s$^{-2}$  below which the gravitational acceleration gets a $\approx 1/r$ behavior. MOND was proposed to explain certain features of the motion of ordinary electromagnetically detectable matter in galaxies and of galaxies in galactic clusters  without resorting to exotic forms of still undetected dark matter.
  \item ${f(R)}$ models \cite{super}. These theories come about by a straightforward generalization
of the Lagrangian in the Einstein-Hilbert action in which  the Ricci scalar $R$
is replaced by a general analytical function $f(R)$ of $R$. They have mainly been used in cosmological and astrophysics scenarios \cite{capoz1}.
  \item {Curvature Invariants} models. They encompass inverse powers \cite{Nav1,Nav2} and logarithm \cite{Nav3} of  some curvature invariants in the Einstein-Hilbert action and have been used for tackling the dark energy-dark matter problems.
  \item {Yukawa}-like models.  There are many theoretical frameworks yielding such a modification of the Newtonian inverse-square law \cite{Kra,Adel,megaberto}. Models encompassing Yukawa-type extra-accelerations have been used for a variety of applications ranging from solar system effects like the Pioneer anomaly \cite{MOG} to astrophysical and cosmological scenarios \cite{Koch,Ame,Sea,je,Shi,Ser06}.
  \item {Hooke}-like models. With such a definition we mean models of gravity introducing an additional term proportional to the distance $r$. An important case is given by the Schwarzschild-de Sitter spacetime \cite{desi} which yields a correction to the Newtonian inverse-square law  proportional to $\Lambda r$ \cite{Rind}, where $\Lambda$ is an uniform cosmological constant. Another example is given by the extra-acceleration proposed by Jaekel and Reynaud \cite{Jekel1} to explain the Pioneer anomaly.
  \item {Pioneer}-like models. The simplest one consists of postulating a constant and uniform acceleration radially directed towards the Sun and having the same magnitude of $A_{\rm Pio}$ existing in the outer regions of the solar system at heliocentric distances $r\geq 20$ AU. Other forms have been postulated for it, both distance-dependent \cite{Jekel1,mofPio} and velocity-dependent \cite{Jekel2,Jekel3,Sta08}.
\end{itemize}
Other models of modified gravity that we will not consider here are the Einstein-Aether theory \cite{Mat}, Tensor-Vector-Scalar (TeVeS) \cite{teves} and braneworld gravity \cite{brane}.

The  only motivations for the aforementioned models are just the phenomena themselves for which they have been introduced.
Thus, such models must be put on the test in different scenarios by devising  independent observational checks. In particular, they  must not exhibit discrepancies with the well-tested standard laws of gravity in local astronomical  systems like, e.g., our solar system and the double pulsar  PSR J0737-3039 \cite{psr1,psr2}.
\section{Our Method for Testing Modified Models of Gravity in Astronomical Scenarios}
In general, a given Long Range Modified Model of Gravity (LRMMG) yields {predictions} ${\mathcal{P}}$ for certain {observable}  effects ${\mathcal{O}}$ of the form \eqi{\mathcal{P} = Kg},\eqf where $K\rightarrow 0$ implies no modifications of gravity, and ${g}$ is a function of the geometrical configuration of the system adopted { characteristic of the LRMMG considered}; the fact that the
LRMMG parameter $K$ enters as a multiplicative factor in the predicted effects $\mathcal{P}$ will be very important for us, as we will see below.
For example,  for the solar system's planetary longitudes of the {perihelia} ${\varpi}$, it turns out \eqi{g=g(a,e)}\eqf with $a$ semimajor axis and $e$  eccentricity of the planetary orbit considered.
{Corrections} ${\Delta\dot\varpi}$ to the usual Newtonian/Einsteinian perihelion precessions have been recently estimated for { several} planets of the solar system by {E.V. Pitjeva} \cite{Pit05,Pit,Pit08} by fitting 100 yr of observations of several kinds with the force models of various versions of the {EPM} ephemerides (EPM2004 \cite{PitSS} EPM2006 \cite{Pit}, EPM2008 \cite{Pit08}). Since they do { not} include  any LRMMGs, the corrections $\Delta\dot\varpi$, {by construction}, account for them, so that they will be {our} ${\mathcal{O}}$: they are listed in Table \ref{chebolas}.
\begin{table}
\caption{Corrections ${\Delta\dot\varpi}$, in ${ 10^{-4} }$ {arcsec cy}${^{-1}}$ ,  to the standard Newton/Einstein perihelion precessions
of the {inner} planets estimated by {E.V. Pitjeva} with the {EPM2004} \protect{\cite{Pit05}} (Mercury, Earth, Mars) and EPM2006 \cite{Pit} (Venus) ephemerides. The result for Venus has been obtained by recently processing radiometric
data from Magellan spacecraft (E.V. Pitjeva, personal communication, 2008). The errors in {square brackets} are  the { formal} ones: the other ones have been {re-scaled} by Pitjeva to get { realistic} estimates.\label{chebolas}
}
\smallskip
\centering
\begin{tabular}{@{}cccc@{}}
\noalign{\smallskip}\hline\noalign{\smallskip}
 {Mercury} & {Venus} & {Earth} & {Mars}  \\
 \noalign{\smallskip}\hline\noalign{\smallskip}
 $-36\pm 50\ [42]$ & $-4\pm 5\ [1]$ & $-2\pm 4\ [1]$ & $1\pm 5\ [1]$\\
\noalign{\smallskip}\hline\noalign{\smallskip}
\end{tabular}
\end{table}
 By {directly} comparing  ${\Delta\dot\varpi}$ to the predicted anomalous perihelion precession ${\mathcal{P}=K g(a,e)}$  for {each planet separately}   allows to put {upper bounds} on ${|K|}$ since $\Delta\dot\varpi$ are compatible with zero, according to Table \ref{chebolas}.  This approach is good for test the hypothesis ${K=0}$.
The hypothesis ${K\neq 0}$ can be tested by taking the {ratios} \eqi{\Pi_{\rm AB}}\equiv \rp{\Delta\dot\varpi_{\rm A}}{\Delta\dot\varpi_{\rm B}}\eqf {of} ${\Delta\dot\varpi}$ for {different pairs} of planets A and B, and comparing them to the {predicted ratios}
\eqi{ \xi_{\rm AB} \equiv\rp{\mathcal{P}_{\rm A}}{\mathcal{P}_{\rm B}}=\rp{g(a_{\rm A},e_{\rm A})}{g(a_{\rm B},e_{\rm B})}}:\eqf
$\xi_{\rm AB}$, by construction, does {not} explicitly depend on ${K}$, but it still retains a functional dependence on $a$ and $e$ { typical of the LRMMG} considered. If \eqi{\Pi_{\rm AB}\neq \xi_{\rm AB}}\eqf within the observational errors, i.e. if \eqi{\Psi_{\rm AB}\equiv\left|\Pi_{\rm AB}-\xi_{\rm AB}\right|\neq 0},\eqf the LRMMG considered is {severely challenged}. Of course, the uncertainty in the ratios $\Pi_{\rm AB}$ has to be evaluated in a realistic and conservative way to reduce the risk of unsound conclusions; see the discussion in Section \ref{finale}. We quote the ratios $\Pi_{\rm AB}$ for all the pairs of the inner planets in Table \ref{chebolas2}.
 \begin{table}
\caption{Ratios $\Pi_{\rm AB}$ of the estimated corrections of the perihelia $\Delta\dot\varpi$ for the pairs of planets A B. The uncertainties $\delta\Pi_{\rm AB}$ have been evaluated by using the realistic errors for the individual corrections of Table \protect\ref{chebolas}.\label{chebolas2}}
\smallskip
\centering
\begin{tabular}{@{}llrr@{}}
\noalign{\smallskip}\hline\noalign{\smallskip}
 A & B  & $\Pi_{\rm AB}$ & $\delta\Pi_{\rm AB}$\\
 \noalign{\smallskip}\hline\noalign{\smallskip}
Mercury & Venus & $9$ & 24 \\
Venus & Mercury & $0.1$ & 0.3 \\
 Mercury & Earth & $18$ & 61 \\
 Earth & Mercury& $0.06$ & 0.19 \\
 Mercury & Mars & $-36$ & 230 \\
 Mars & Mercury& $-0.03$ & 0.18 \\
Venus & Earth & $2$ & 6.5 \\
 Earth & Venus& $0.5$ & 1.6 \\
 Venus & Mars & $-4$ & 25 \\
 Mars & Venus & $-0.2$ & 1.6 \\
 Earth & Mars & $-2$ & 14 \\
 Mars & Earth & $-0.5$ & 3.5 \\
\noalign{\smallskip}\hline\noalign{\smallskip}
\end{tabular}
\end{table}

An analogous approach can be followed with the {double pulsar} PSR J0737-3039. Now, one observable is ${  \Delta\dot\omega  }$, i.e. the difference between the
{phenomenologically determined} periastron precession ${  \dot\omega_{\rm meas}  }$ and the {computed} rate ${  \dot\omega_{\rm 1PN}  }$ for the general relativistic 1PN rate (it is analogous to the well-known Mercury's perihelion precession by Einstein). Another observable is ${  \Delta P }$, i.e. the difference between the {phenomenologically determined} orbital period
${ P_{\rm b}  }$ and the {computed} Keplerian one ${  P^{\rm Kep}  }$. Thus, the {observation-based} ratio \eqi{\mathcal{R}}\equiv \rp{\Delta\dot\omega}{\Delta P}\eqf can be compared with the  {predictions} $\mathcal{P}$ for the same ratio by various {LRMMGs}; again, if they are {equal} within the errors, LRRMMG { passes} the test, {otherwise} it is { challenged}.
\section{Results from Local Systems}
\subsection{Dvali-Gabadadze-Porrati (DGP) Braneworld Model}
The { DGP braneworld} model \cite{DGP}, put forth to explain the cosmic acceleration without resorting to dark energy, predicts an anomalous {perihelion precession} independent of the semimajor axis  \cite{lue,DV,IorDGP}
\eqi{ \dot\varpi_{\rm DGP}=\mp\frac{3c}{8r_0}=\mp 0.0005\ {\rm arcsec\ cy^{-1}}}, \ r_0\approx 5\ {\rm Gpc},\eqf where the $\mp$ sign are related to the two different cosmological
branches of the model. Thus, \eqi{\xi_{\rm AB}=1}\eqf for it.
By linearly adding the absolute values of the uncertainties in $\Delta\dot\varpi$, it turns out  for the pairs A=Mars, B=Mercury and A=Earth, B=Mercury   \cite{IorAHEP}
\begin{eqnarray}
  \Psi_{{\rm Mars\ Mercury}} &=& {1.0\pm 0.2},\  {\rm ruled\ out\ at}\ {5\sigma}\ {\rm level}\\\nonumber\\
   \Psi_{{\rm Earth\ Mercury}} &=& {0.9\pm 0.2},\ {\rm ruled\ out\ at}\ {4.5\sigma}\ {\rm level}
\end{eqnarray}

In the case of the {double pulsar} \cite{Iorpul}, the {predicted} ratio ${\mathcal{R}_{\rm DGP}}$ between $\dot\omega_{\rm DGP}$ and $P_{\rm DGP}$ is
\eqi{ \mathcal{R}_{\rm DGP} = (1.4\pm 0.1)\times 10^{-7}\ {\rm s}^{-2} }.\eqf
Instead, the {observed} ratio ${\mathcal{R}_{\rm obs}}$ between $\Delta\dot\omega\equiv \dot\omega_{\rm meas}-\dot\omega_{\rm 1PN}$ and $\Delta P=P_{\rm b}-P^{\rm Kep}$ is
\eqi{ \mathcal{R}_{\rm meas} = (0.3\pm 4)\times 10^{-11}\ {\rm s}^{-2} }.\eqf
They are {not} compatible at ${14\sigma}$ level.
\subsection{MOdified Gravity (MOG)}
{MOG} \cite{MOG} predicts a {Yukawa-type} extra-acceleration  \cite{mog1,mog2}
\eqi{ A_{\rm MOG} = -\frac{GM}{r^2}\alpha\left[1-\left(1+\mu r\right)\exp{(-\mu r)}\right]},\eqf
where $\alpha$ and $\mu$ have been fitted to a set of astrophysical galactic  data \cite{mog2} in the
framework of the searches for an explanation of the flat
rotation curves of galaxies without resorting to dark matter. For a Sun-planet system we have a perihelion extra-rate  (${1/\mu = 33,000}$ {AU}) \cite{IorMOG}
\eqi{  \dot\varpi_{\rm MOG}\approx -\alpha\mu^2\sqrt{GMa(1-e^2)}  }.\eqf
The inner planets yield  \cite{IorMOG}
\begin{eqnarray}
  \Psi_{{\rm Venus\ Mercury}} &=& {1.3\pm 0.3},\ {\rm ruled\ out\ at\ } {4\sigma}\ {\rm level}\\\nonumber\\
  \Psi_{{\rm Earth\ Mercury}} &=& {1.6\pm 0.2},\ {\rm ruled\ out\ at\ } {8\sigma}\ {\rm level}\\\nonumber\\
  \Psi_{{\rm Mars\ Mercury}} &=& {2\pm 0.2},\ {\rm ruled\ out\ at\ } {10\sigma}\ {\rm level}
\end{eqnarray}

\subsection{MOdified Newtonian Dynamics (MOND) }
MOND \cite{Mil83a,Mil83b,Mil83c} predicts {acceleration-dependent} modifications  of the Newtonian inverse-square law  to explain the {almost flat galactic rotation curves}
\eqi{  A=\frac{A_{\rm Newton}}{\mu(x)},\ x \equiv \frac{A}{A_0},}\eqf  where galaxies data fitting yields \cite{Bege} ${A_0=1.2\times 10^{-10}\ {\rm m\ s}^{-2}  }$ and ${\mu(x)}$ is an {interpolating function} whose most widely used forms  are
\eqi{ \mu = \frac{x}{\sqrt{1+x^2} }\ \protect{\cite{joint}},\  \mu=\frac{x}{1+x}\ \protect{\cite{fam}},}\eqf with \eqi  \mu\rightarrow 1,\ x\gg 1,\eqf and \eqi\mu\rightarrow x,\ x\ll 1.\eqf
{solar system} observations are, strictly speaking, tests of the {form} of the interpolating function in the {large acceleration limit}  (${x\gg 1}$).

The anomalous {perihelion precession} predicted by MOND is, in the large acceleration limit (${x\gg 1}$), \cite{Mil83a,Tal,Ser}
\eqi{  \dot\varpi_{\rm MOND} = -k_0 n\left(\frac{a}{r_{\rm MOND}}\right)^{2h} h,\ n=\sqrt{\frac{GM}{a^3}},\ r_{\rm MOND}=\sqrt{\frac{GM}{A_0}}  };\eqf
${k_0=1,\ h=1}$ correspond to \eqi{\mu = \frac{x}{1+x} };\eqf ${ k_0=1/2,\ h=2}$ correspond to \eqi{ \mu=\frac{x}{\sqrt{1+x^2} }}.\eqf
It turns out \cite{IorMOND} that the {ratios} ${\Pi_{\rm AB}}$ of the perihelia, {independent} by construction of ${k_0}$, for the pairs {A=Mars, B=Mercury} and {A=Earth, B=Mercury} {rule out} the MOND perihelion precessions for $ {1\leq h \leq 2}$ at {several sigma} level.
%
\subsection{Extended $f(R)$ theories}
Extended theories of gravity with \cite{capoz2} \eqi{  f(R)=f_0 R^k  },\eqf  where ${R}$ is the {Ricci scalar}, yield  {  power-law corrections} to the Newton's law
\eqi{  A_{f(R)} = \frac{(\beta - 1)GM}{r^{\beta}_c}r^{\beta - 2}  }\eqf which, for ${\beta = 0.817}$, obtained by fitting the SNeIa Hubble diagram, yields good results in fitting some galactic rotation curves; $\beta$ is related to the exponent $k$ of the Ricci scalar $R$.
The induced perihelion precession is \cite{IorRug}
\eqi{   \dot\omega_{f(R)}=\frac{(\beta - 1)\sqrt{GM}}{2\pi r_c^{\beta}}a^{\beta - 3/2} G(e,\beta)  },\eqf where $G(e,\beta)$ is a function of the eccentricity and $\beta$.
The resulting $\Psi_{\rm AB}$ are {not} compatible with zero for {many} pairs A and B of inner and outer planets at {several sigma} level \cite{IorRug}.
\subsection{Inverse powers and logarithm of some curvature invariants models}
{ Inverse powers} of  {curvature invariants} in the action lead to  \cite{Nav1,Nav2}
\eqi{ A_{\rm inv.\ pow.} = -\frac{\alpha GM(6k+3)}{r_c^{6k+4}} r^{6k+2}  }.\eqf For ${k=1}$ and the {Sun}, ${r_c=10}$ {pc}, so that
 \eqi{  \dot\varpi_{k=1}\approx -\frac{45\alpha}{r_c^{10}}\sqrt{GM a^{17}(1-e^2)} }\eqf Thus, the {inner} planets yield  \cite{IorAHEP}
\begin{eqnarray}
  \Psi_{{\rm Mars\ Mercury}} &=& {10^5\pm 0.1},\ {\rm ruled\ out\ at\ } {10^6 \sigma}\ {\rm level}\\ \nonumber \\
  \Psi_{{\rm Mars\ Earth}} &=& {38\pm 3.5},\ {\rm ruled\ out\ at\ } {11\sigma}\ {\rm level}  \\ \nonumber \\
  \Psi_{{\rm Earth\ Mercury}} &=& {10^3\pm 0.2},\ {\rm ruled\ out\ at\ } {10^4 \sigma}\ {\rm level}
\end{eqnarray}

{Logarithm} of some {curvature invariants} in the action, able to modify gravity at {MOND scales} so to {jointly} treat {Dark Matter} and {Dark Energy}, induces  \cite{Nav3}
\eqi{  A_{\rm Log}\propto \frac{r^2}{r_c^4},\ r_c = 0.04\ {\rm pc}  },\eqf
so that
\eqi{  \dot\varpi_{\rm Log}\propto \alpha\frac{\sqrt{GMa^5(1-e^2)}}{r_c^4}   }\eqf
The pairs {A=Mars, B=Mercury} and {A=Earth, B=Mercury} yield  \cite{IorAHEP}
\begin{eqnarray}
  \Psi_{{\rm Mars\ Mercury}} &= &{30.7\pm 0.1},\ {\rm ruled\ out\ at\ } {31 \sigma}\ {\rm level}\\ \nonumber \\
  \Psi_{{\rm Earth\ Mercury}} &=& {10.6\pm 0.2},\ {\rm ruled\ out\ at\ } {53 \sigma}\ {\rm level}
\end{eqnarray}
\subsection{Yukawa-like theories }
Many theoretical frameworks  \cite{Kra,Adel,megaberto} yield a Yukawa-type acceleration
\eqi{  A_{\rm Yukawa} = -\frac{GM\alpha}{r^2}\left(1+\frac{r}{\lambda}\right)\exp\left(-\frac{r}{\lambda}\right)  }.\eqf
For ${\lambda\gtrsim ae}$, the induced  perihelion rate is    \cite{Iyuk}
\eqi{  \dot\varpi_{\rm Yuk}\approx \frac{\alpha\sqrt{GMa}}{2\lambda^2}\exp\left(-\frac{a}{\lambda}\right)   },\eqf so that for a pair of planet A and B
\eqi{  \lambda = \frac{a_{\rm B} - a_{\rm A}}{\ln\left(\sqrt{\frac{a_{\rm B}}{a_{\rm A}}}\Pi_{\rm AB}\right)}  }.\eqf

The pair {A=Earth, B=Mercury} allows to obtain   \cite{Iyuk}
\eqi{  \lambda = 0.182\pm 0.183\ {\rm AU}  }.\eqf
Solving for ${\alpha}$ and using $\Delta\dot\varpi$ and $a$ for { Mars } yields  \cite{Iyuk}
\eqi{ \alpha = \frac{2\lambda^2\Delta\dot\varpi}{\sqrt{GMa}}\exp\left(\frac{a}{\lambda}\right)= (0.2\pm 1)\times 10^{-9}   },\eqf
in which we have used the value for ${\lambda}$ obtained from {Earth} and {Mercury}.
\subsection{Hooke-type theories}
A Hooke-like acceleration is induced, e.g., by a uniform {cosmological constant} ${\Lambda}$ in the {Schwarzschild-De Sitter} spacetime \cite{Rind}
\eqi{  A_{\Lambda}=\frac{1}{3}\Lambda c^2 r  }.\eqf
The induced perihelion shift is  \cite{Kerr}
\eqi{ \dot\varpi_{\Lambda} = \frac{1}{2}\left(\frac{\Lambda c^2}{n}\right)\sqrt{1-e^2},\ n=\sqrt{\frac{GM}{a^3}}   }.\eqf
Some pairs of planets yield \cite{IorH}
\begin{eqnarray}
  \Psi_{{\rm Mars\ Mercury}} &=& {7.8\pm 0.2},\ {\rm ruled\ out\ at\ } {39 \sigma}\ {\rm level},\\ \nonumber\\
  \Psi_{{\rm Earth\ Mercury}} &=& {4.1\pm 0.2},\ {\rm ruled\ out\ at\ } {20 \sigma}\ {\rm level},\\ \nonumber\\
  \Psi_{{\rm Jupiter\ Mercury}} &=& {51\pm 12},\ {\rm ruled\ out\ at\ } {4 \sigma}\ {\rm level}
\end{eqnarray}

In the case of the {double pulsar}, the {predicted} ratio ${\mathcal{R}_{\Lambda}}$ between $\dot\omega_{\Lambda}$ and $P_{\Lambda}$ is  \cite{Iorpul}
\eqi{ \mathcal{R}_{\Lambda} = (3.4\pm 0.3)\times 10^{-8}\ {\rm s}^{-2} }.\eqf
Instead, the {observed} ratio ${\mathcal{R}_{\rm obs}}$ between $\Delta\dot\omega\equiv \dot\omega_{\rm meas}-\dot\omega_{\rm 1PN}$ and $\Delta P=P_{\rm b}-P^{\rm Kep}$  is   \cite{Iorpul}
\eqi { \mathcal{R}_{\rm meas} = (0.3\pm 4)\times 10^{-11}\ {\rm s}^{-2} }.\eqf
They are {not} compatible at ${11\sigma}$ level.

\subsection{The Pioneer anomaly}
If the {Pioneer Anomaly} \cite{Pio1,Pio2} was of {gravitational origin}, the exotic force causing it {should also act on the planets} of the Solar System, at least on those  moving in the spatial regions in which it manifested itself in  its presently known form. A {constant and uniform radial} acceleration with the same magnitude of that causing the Pioneer Anomaly would induce orbital effects {too large} \cite{IorGiu} to have escaped so far detection, as shown by the ${\Delta\dot\varpi}$ estimated with the {EPM2006} ephemerides \cite{Pit} for {Jupiter, Saturn and Uranus} \cite{IorMGM,IorJgP}, and from the {residuals} of right ascension $\alpha$ and declination $\delta$ computed with the {JPL} ephemerides \cite{IorFoP}. Also some recently proposed {velocity-dependent} exotic forces \cite{Jekel2,Jekel3,Sta08} are {ruled out} by planetary observations \cite{IorFoP,IorIJ}.  For other works dealing with the same problem, see Refs.~\cite{Page06,Tangen,Wal,Page09}.
\subsection{The anomalous perihelion precession of Saturn}
{E.V. Pitjeva} has recently fitted almost one century of planetary data of all kinds, including also  {3 years of radiotechnical data from Cassini}, with the {EPM2008} ephemerides \cite{Pit08}. She estimated (E.V. Pitjeva, private communication, 2008) a statistically significant non-zero correction
\eqi{  \Delta\dot\varpi_{\rm Saturn} = -0.006\pm 0.002\ {\rm arcsec\ cy^{-1}}  },\eqf while the {formal} error of the fit is {0.0007   arcsec cy$^{-1}$ }. However, applying the standard re-scaling by a factor 10 of the optical only observations there is also, in principle,  the possibility that the uncertainty can be as large as 0.007 arcsec cy$^{-1}$ (E.V. Pitjeva, private communication, 2008).
Previous results obtained with the EPM2006 ephemerides \cite{Pit}, which did not include Cassini data, yielded
\eqi \Delta\dot\varpi_{\rm Saturn} = -0.92\pm 0.29\ {\rm arcsec\ cy^{-1}}\ ({\rm formal\ error}).\eqf
Is it really a genuine physical effect needing explanation, or is it some artifact of the data reduction procedure? {None} of the exotic models examined so far {is able to accommodate  it} \cite{IorSat}.
\subsection{The general relativistic Lense-Thirring effect}
Until now we have only dealt with putative modifications of the standard Newtonian/Einsteinian laws of gravity. In fact, the estimated corrections $\Delta\dot\varpi$ to the usual rates of the perihelia account, by construction, also for a standard general relativistic effect which has not been included in the dynamical force models of the EPM ephemerides used to determine them: it is the gravitomagentic Lense-Thirring effect \cite{LT} consisting of secular precessions of the form
\eqi\dot\varpi_{\rm LT}=\rp{2GS(1-3\cos I)}{c^2 a^3(1-e^2)^{3/2}},\eqf where $S$ is the Sun's angular momentum and $I$ is the inclination of the planetary orbital plane to the Sun's equator which is quite small for all the inner planets. The Lense-Thirring precessions for the inner planets are of the order of $10^{-3}-10^{-4}$ arcsec cy$^{-1}$, and their direct measurability has been discussed in, e.g., Ref.~\cite{POS}. It is interesting to check if the unmodelled Lense-Thirring precessions pass the test of the ratio of the perihelia, with
\eqi\xi^{\rm LT}_{\rm AB}\approx  \rp{a_{\rm B}^3 (1-e^2_{\rm B})^{3/2}}{a_{\rm A}^3 (1-e^2_{\rm A})^{3/2}}.\eqf By comparing Table \ref{chebolas2} with Table \ref{chebolas3}, it can be seen that
$\Psi_{\rm AB}$ is compatible with zero for all the pairs A B of inner planets, contrary to all the models of modified gravity examined so far.
 \begin{table}
\caption{Ratios $\xi^{\rm LT}_{\rm AB}$ of the predicted Lense-Thirring precessions for the pairs of planets A B.\label{chebolas3}}
\smallskip
\centering
\begin{tabular}{@{}lll@{}}
\noalign{\smallskip}\hline\noalign{\smallskip}
 A & B  & $\xi^{\rm LT}_{\rm AB}$ \\
 \noalign{\smallskip}\hline\noalign{\smallskip}
Mercury & Venus & $7$  \\
Venus & Mercury & $0.1$  \\
 Mercury & Earth & $18$  \\
 Earth & Mercury& $0.05$  \\
 Mercury & Mars & $64$  \\
 Mars & Mercury& $0.01$  \\
Venus & Earth & $3$  \\
 Earth & Venus& $0.4$  \\
 Venus & Mars & $9$  \\
 Mars & Venus & $0.1$  \\
 Earth & Mars & $3.5$  \\
 Mars & Earth & $0.3$  \\
\noalign{\smallskip}\hline\noalign{\smallskip}
\end{tabular}
\end{table}

\section{Discussion and conclusions}\lb{finale}
\subsection{Some technical aspects}
{If}
and {when} {other} teams of astronomers will {independently} estimate their own corrections $\Delta\dot\varpi$ with {different} ephemerides, it will be possible to fruitfully repeat all the tests presented here. In doing them we {always} used the formal errors in $\Delta\dot\varpi$ {re-scaled} by a factor ${\approx 2-5}$ for the { inner} planets and up to {10} times for the {outer} planets for which mainly optical data have been used. Moreover, in view of the {correlations} among the estimated  $\Delta\dot\varpi$, we always {linearly} propagated their {errors}, instead of summing them in quadrature, by getting
\eqi\delta\Pi_{\rm AB}\leq\left|\Pi_{\rm AB}\right|\left( \rp{\delta\Delta\dot\varpi_{\rm A}}{|\Delta\dot\varpi_{\rm A}|}+ \rp{\delta\Delta\dot\varpi_{\rm B}}{|\Delta\dot\varpi_{\rm B}|}\right).\eqf For example, the correlations
between the corrections $\Delta\dot\varpi$ of Mercury and the Earth are as large as $20\%$ (Pitjeva, private communication, 2005).
Concerning the uncertainties in $\xi_{\rm AB}$, in principle, they should have been computed by propagating the errors in the semimajor axes $a$ and the eccentricities $e$ of the planets A and B entering them for each model considered. However,  they are quite negligible because the relative (formal) uncertainties in the semimajor axes and eccentricities of the inner planets are all of the order of $10^{-12}$ and $10^{-9}-10^{-11}$, respectively, according to Table 3 of Ref.~\cite{Pit}.
Processing { more ranging data from Cassini} will help in clarifying if {the perihelion precession of Saturn} can really be considered as a {genuine physical effect}.

\subsection{Conclusions}
\begin{itemize}
  \item{ All} the long-range modified models of gravity examined here are {severely challenged} either by {the ratios of the perihelia} of different pairs of solar system's planets or by the {double pulsar} combined data for the periastron and the orbital period. Both such kinds of ratios cancel out the small multiplicative parameters which directly account for the various exotic models considered, but they still retain a pattern which is characteristic of the model tested through the orbital parameters of the systems used. Only the general relativistic Lense-Thirring effect, not modelled in the EPM ephemerides and, thus, accounted for, in principle, by the estimated corrections $\Delta\dot\varpi$, passes the test of the ratio of the perihelia.
 \item{None} of the exotic models examined {is able to explain the anomalous perihelion precession of Saturn}.
  The analysis of {more  Cassini data} will help in clarifying if it is really incompatible with zero at some statistically significant level, thus requiring an explanation in terms of some physical phenomena, or if it is some artifact of the data reduction procedure.
\end{itemize}

\acknowledgments{I gratefully thank the organizers and the entire staff of this prestigious and high-quality international school for their kind invitation, their exquisite hospitality and  the financial support received.}

\end{document}